\documentclass[a4paper,dvips,12pt]{article}
\usepackage{epsfig,graphics,graphicx}
\begin{document}
\pagestyle{myheadings}
\markright{\it P0193}
\vskip.5in
\begin{center}
\vskip.4in {\Large\bf On the Casimir energy of a massive scalar field in  positive
curvature space }
\vskip.3in
E Elizalde\footnote{Email: \tt elizalde@ieec.fcr.es }\\
Consejo Superior de Investigaciones Cient\'{\i}ficas \\
 Institut d'Estudis Espacials de Catalunya (IEEC/CSIC) \\
 Edifici Nexus, Gran Capit\`{a} 2-4, 08034 Barcelona, Spain;\\
 Departament d'Estructura i Constituents de la Mat\`{e}ria \\
 Facultat de F\'{\i}sica, Universitat de Barcelona \\
  Diagonal 647, 08028 Barcelona, Spain \\

A C Tort\footnote{Email: \tt tort@if.ufrj.br}\\
Departamento de F\'{\i}sica Te\'{o}rica - Instituto de F\'{\i}sica\\
Universidade Federal do Rio de Janeiro\\
Caixa Postal 68.528; CEP 21941-972 Rio de Janeiro, Brazil

\end{center}
\vskip.2in
\begin{abstract}
We re-evaluate the zero point Casimir energy for the case of a massive scalar field in $\mathbf{R}^{1}\times\mathbf{S}^{3}$ space, allowing also for deviations from
the standard conformal value $\xi =1/6$, by means of zero temperature zeta function techniques. We show that for the problem at hand this approach is equivalent to the high
temperature regularisation of the vacuum energy.
\end{abstract}
%% \section{Introduction}
Recently the vacuum and finite temperature energies for
massless and  massive scalar fields in positive curvature spaces,
as $\mathbf{S}^3$, were considered, with special emphasis being
put on the analysis of  entropy bounds corresponding to those
cases [see Refs. \cite{Breviketal2002} and
\cite{Elizalde&Tort2003}, respectively]. Specifically, in Ref.
\cite{Elizalde&Tort2003}, dealing with the massive case and
arbitrary coupling, the vacuum (Casimir) energy and the finite
temperature energy were obtained through the application of a
generalised zeta function technique \cite{Elizalde94}, which
provides a neat formal separation of the logarithm of the
partition function into the non-thermal and thermal sectors. This
approach holds generally and, in particular, it does for
$\textbf{S}^1\times \textbf{S}^d$ spaces. For the special case
$d=3$, the result for the logarithm of the partition function can
be combined with the Abel-Plana rescaled sum formula
\cite{Erdelyietal1953}, leading to \cite{Elizalde&Tort2003}
\begin{eqnarray}\label{logZ}
\log Z\left(\beta \right)&=&-\frac{\beta }{2r}\sum _{n=1}^{\infty }n^{2}
\left(n^{2}+\mu _{eff}^{2}\right)^{1/2}+\frac{r\mu _{eff}^{2}}{\beta }
\sum _{n=1}^{\infty }\frac{1}{n^{2}}K_{2}\left(m_{eff}\beta n\right)
 \nonumber \\ &+&\frac{\mu _{eff}^{2}\beta }{\left(2\pi \right)^{2}r}\,
\sum _{n=1}^{\infty }\frac{1}{n^{2}}K_{2}\left(2\pi n\mu
_{eff}\right) -\frac{\mu _{eff}^{3}\beta }{2\pi r}\sum
_{n=1}^{\infty }\frac{1}{n}K_{3} \left(2\pi n\mu _{eff}\right).
\end{eqnarray}
From this result the present authors conjectured that the
\emph{renormalised} vacuum energy could be inferred as
\begin{equation}\label{Representation1}
E_0=-\frac{\mu^2_{eff}}{\left(2\pi\right)^2r}\sum_{n=1}^\infty
\frac{1}{n^2}K_2\left(2\pi
n\mu_{eff}\right)+\frac{\mu^3_{eff}}{2\pi r}
\sum_{n=1}^\infty\frac{1}{n}K_3\left(2\pi n\mu_{eff}\right),
\end{equation}
where $r$ is the radius of $\mathbf{S}^3$, $\beta$ is the
reciprocal of the temperature, and the parameter $\mu_{eff}$, to
be defined below, plays the role of an `effective mass'. The
conjecture stems from the fact that in Eq.~(\ref{logZ}), the terms
linear in $\beta$ give rise to temperature independent terms when
we calculate basic thermodynamics quantities such as the free
energy and the energy. Also, in the very-high temperature limit,
it is physically plausible to expect  the Stefan-Boltzmann term
---contained in the second term in Eq.~(\ref{logZ})--- to be
the only surviving one. 

here we show that the above-mentioned conjecture
holds true and that the standard zero-temperature zeta
function regularisation procedure (as prescribed in
\cite{Elizalde94}, for instance)  works well, yielding the
correct vacuum energy for the non-conformal case.

The vacuum energy for a massive scalar field in $\mathbf{S}^3$ and
arbitrary conformal parameter $\xi$ is given by
\begin{equation}
E_0=\frac{1}{2r}\sum_{\ell=0}^\infty D_{\ell}M_{\ell},
\end{equation}
where
\begin{equation}
M_{\ell }^{2}:=\left(\ell +1\right)^{2}+\mu _{eff}^{2}
\end{equation}
and the dimensionless parameter $\mu _{eff}^{2}$ is defined as
\begin{equation}\label{effective mass}
\mu _{eff}^{2}=\mu ^{2}+\chi -1,
\end{equation}
where $\mu:=mr$, $m$ being the mass of an elementary excitation of
the scalar field, $r$ the radius of $\mathbf{S}^3$, and
$\chi:=\xi{\cal R}r^2$; ${\cal R}$ is the Ricci curvature scalar.
Notice that $\mu
_{eff}^{2}$ and $\chi$ are real and $\mu ^{2}\geq 0$.
The degeneracy factor is $D_{\ell }=\left(\ell +1\right)^{2}$.
Hence, for a massive scalar field the unregularised Casimir energy is 
given by
\begin{equation}\label{zero-point energy}
E_{0}\left(\mu _{eff}^{2}\right)=\frac{1}{2r}\sum _{\ell
=0}^{\infty } \left(\ell +1\right)^{2}\sqrt{\left(\ell
+1\right)^{2}+\mu _{eff}^{2}}.
\end{equation}
The standard conformal case corresponds to the
values $\mu ^{2}=0$ and $\chi =1$ ($\xi =1/6$, ${\cal R}=6/r^2$)
and, using the zeta function prescription, we
obtain
\begin{equation}
E_{0}\left(\mu ^{2}=0,\chi =1\right)=\frac{1}{2r}\zeta \left(-3\right)=
\frac{1}{240r}.
\end{equation}
The problem now is to find an analytical continuation in
terms of the zeta function for the more difficult series in
Eq.~(\ref{zero-point energy}).

Starting from Eq.~(\ref{zero-point energy}) we can
straightforwardly write:
\begin{equation}
\left. \sum _{\ell=0}^{\infty
}\left(\ell+1\right)^{2}\left[\left(\ell+1\right)^{2}+\mu_{eff}^2
\right]^{-s} \right|_{s=-\frac{1}{2}}=F(-3/2,\mu_{eff}^2) -
\mu_{eff}^2F(-1/2,\mu_{eff}^2), \label{soli1}
\end{equation}
where by definition,
\begin{equation}
F(s,\mu_{eff}^2) :=\sum _{\ell=0}^{\infty }
\left[\left(\ell+1\right)^{2}+\mu_{eff}^2 \right]^{-s}
\end{equation}

The sum over $\ell$ can be
identified as an Epstein series whose analytical continuation is well-known  \cite{Elizalde94,ER89a}.
It follows that the vacuum energy is
\begin{equation}
E_0=-\frac{\mu^4_{eff}}{16r}\Gamma\left(-2\right)+\frac{3\mu^2_{eff}}{4\pi^2
r} \sum_{n=1}^\infty\frac{1}{n^2}K_2\left(2\pi n\mu_{eff}\right)
+\frac{\mu^3_{eff}}{2\pi
r}\sum_{n=1}^\infty\frac{1}{n}K_1\left(2\pi
n\mu_{eff}\right).\label{Sumpri}
\end{equation}
The divergent term in Eqs.~(\ref{Sumpri}) can be taken care of with
the minimal subtraction scheme \cite{Blauetal1988}. However, this
procedure would mean, in our case, to keep a quartic term
in $\mu_{eff}$ that would spoil the behaviour of the vacuum
energy in the classic limit where the vacuum oscillations
must vanish. Therefore, without too much ado we will simply
discard this term. Now, we can further simplify Eq.~(\ref{Sumpri}) and write
\begin{eqnarray}
E_0&=&\frac{\mu^2_{eff}}{4\pi^2 r}\sum_{n=1}^\infty
\frac{1}{n^2}K_2\left(2\pi n\mu_{eff}\right) \nonumber \\
&+&\frac{\mu^3_{eff}}{4\pi
r}\sum_{n=1}^\infty\frac{1}{n}K_1\left(2\pi n
\mu_{eff}\right)+\frac{\mu^3_{eff}}{4\pi
r}\sum_{n=1}^\infty\frac{1}{n} K_3\left(2\pi
n\mu_{eff}\right).\label{Representation2}
\end{eqnarray}
If we now make use of the recursion relation
\[
K_{\nu-1}\left(z\right)-K_{\nu+1}\left(z\right)=-\frac{2\nu}{z}K_\nu
\left(z\right),
\]
we easily obtain Eq.~(\ref{Representation1}),
as conjectured in a former paper \cite{Elizalde&Tort2003}.

If we take the limit $\mu_{eff}^2\to 0$,
Eqs.~(\ref{Representation1}) and (\ref{Representation2}) yield the
well known result $E_0\approx 1/240$. If we make use of
the appropriate small argument expansion of the Bessel functions
of the second kind we obtain
\begin{equation}
E_0\approx \frac{1}{240 r}-\frac{\mu^2_{eff}}{48
r}-\frac{1}{2}
\left[\frac{1}{8r}+\frac{1}{16r}\left(-\frac{3}{2}+2
\gamma_E\right)\right]\mu^4_{eff}.
\end{equation}
If we take the opposite limit, $\mu_{eff}^2\to
\infty$, the vacuum energy given by Eq.~(\ref{Representation1})
---or (\ref{Sumpri}) or (\ref{Representation2})--- behaves in the
way that one would normally expect of constrained zero-point
oscillations of a massive quantum field: the vacuum energy goes to
zero in an exponential way. 

A detailed version of the calculation briefly described here can be found in \cite{Elizalde&Tort2004}.
\section*{Acknowledgments}
A.C.T. acknowledges the financial support of XXIV ENFPC Organizing Comitee that allowed him to be part of the XXIV Brazilian National Meeting on Particles and Fields held in Caxambu MG, Brazil, September 30 -- October 4, 2003. A.C.T. also acknowledges the kind hospitality of the Institut
d'Estudis Espacials de Catalunya (IEEC/CSIC) and the Universitat
de Barcelona, Departament d'Estructura i Constituents de la
Mat\`{e}ria, where the present work began. The investigation
of E.E. has been supported by DGI/SGPI (Spain), project
BFM2000-0810, and by CIRIT (Generalitat de Catalunya), contract
1999SGR-00257.

\end{document}